\documentclass[conference]{IEEEtran}
\usepackage{listings}
\usepackage[numbers]{natbib}
\usepackage{hyperref}
\usepackage[shortlabels, inline]{enumitem}
\usepackage{graphicx}
\usepackage[table]{xcolor}

\title{Presentation: Reviewing KLEE's Sonar-Search Strategy in Context of Greybox Fuzzing \\ {\large \textbf{An extended abstract}}}

\author{\IEEEauthorblockN{Saahil Ognawala, Alexander Pretschner, Thomas Hutzelmann, Eirini Psallida, Ricardo Nales Amato}
    \IEEEauthorblockA{\textit{Technical University of Munich, Germany} \\
        \{firstname.secondname\}@tum.de}
}

\begin{document}

\maketitle
\definecolor{orange}{rgb}{1,0.5,0}



\section{Introduction}\label{sec:introduction}
Automatic test-case generation techniques of symbolic execution and fuzzing are the most widely used methods to discover vulnerabilities in, both, academia and industry. However, both these methods suffer from fundamental drawbacks that stop them from achieving high path coverage that may, consequently, lead to discovering vulnerabilities at the numerical scale of static analysis. In this presentation, we examine systems-under-test (SUTs) at the granularity level of \emph{functions} and postulate that achieving higher function coverage (execution of functions in a program at least once) than, both, symbolic execution and fuzzing may be a necessary condition for discovering more vulnerabilities than both. 

We will start this presentation with the design of a \emph{targeted search strategy} for KLEE, sonar-search, that prioritizes paths leading to a target function, rather than maximizing overall path coverage in the program. Then, we will show that examining SUTs at the level of functions (compositional analysis \cite{ognawala2016macke}) leads to discovering more vulnerabilities than symbolic execution from a single entry point. Using this finding, we will, then, demonstrate a greybox fuzzing method \cite{ognawala2018munch} that can achieve higher function coverage than symbolic execution. Finally, we will present a framework to effectively manage vulnerabilities and assess their severities \cite{severityassessment}. 
\section{Sonar-Search Strategy}\label{sec:sonar-search}
We implemented a targeted search strategy in KLEE to reach entry points of arbitrary functions. 
Since the original implementation of KLEE does not provide such a targeted strategy, we extended it with \emph{sonar-search} in KLEE22 \cite{klee22repo}, our fork of KLEE. We drew inspiration from past works such as \cite{pretschner2001classical,ma2011directed} and adapted them so that we may use LLVM control-flow-graph and call-graph to calculate node distances, instead of relying solely on the source code.

Sonar-search works in the following way -- for every state that can be reached by the symbolic executor, the searcher 
tracks the minimum number of steps, \texttt{minFutureDistance}, that this state is \emph{yet to take} to reach the target function's entry point. 
Then, for every new state during symbolic execution, there may be three possible values for \texttt{minFutureDistance}
\begin{enumerate}
    \item If the target entry point cannot be reached, then the value is $\infty$. 
    \item If the target entry point can be reached, then the value is the maximum of 
    \begin{enumerate}
        \item the number of direct steps (without tracking back in the program stack) to the target entry point, and
        \item \texttt{minFutureDistance} of its direct ancestor in the program stack plus the minimum steps till \texttt{return} statement, so as to reach the direct ancestor. 
    \end{enumerate}
\end{enumerate}
Naturally, then, all the execution states that can never reach the target entry point ($\texttt{minFutureDistance}=\infty$) are removed from the queue of the states to be explored. 

Finally, so as to not stop exploring more paths in the program, KLEE22 switches to KLEE's default search strategy for all states that follow the target function's entry point.
\section{Compositional Symbolic Execution with Macke}\label{sec:macke}
Sonar-search strategy is a building block of Macke \cite{ognawala2016macke}, our compositional analysis tool that employs parallel and targeted symbolic execution. The goal of compositional analysis is to find more vulnerabilities than forward symbolic execution, while, at the same time, keeping the number of false-positives less than static analysis. 

Macke starts by isolating functions to prepare them for parallel symbolic execution. All functions get individual entry-points at the LLVM level, by analyzing the number and types of formal arguments (primitive datatype or single pointers only). After the functions are isolated, Macke symbolically executes every function for a short amount of time. This step leads to higher path coverage than KLEE for functions lying low in the control-flow-graph of a program, for the simple reason that Macke removes the guard conditions (constraints) to trigger the functions. As a result, Macke can find more vulnerabilities in the program than KLEE. 

However, the vulnerabilities found by Macke in isolated functions may not be ``real'' vulnerabilities if there is no opportunity to exploit them using a usable entry-point to the program. To estimate the impact of the discovered vulnerabilities, Macke performs compositional symbolic execution by pruning those paths from higher level functions that do not lead to discovered vulnerabilities. Sonar-search is the strategy of choice for this path-pruning exercise. Macke, in this step, replaces function bodies (using LLVM Opt) by simple assertions that compare incoming function parameters with exploits generated by KLEE22 for isolated functions. Then, Macke uses sonar-search strategy, with vulnerable functions as targets, to confirm whether those vulnerabilities can be exploited from higher level functions.

Our evaluation results on 15 open-source programs, in \cite{ognawala2016macke} and another unpublished study, show that there is an increase in instruction coverage with Macke, as compared to KLEE, trivially due to adding multiple entry-points. Additionally, we were able to generate 108 exploits for vulnerabilities in different versions of these 15 programs, that could be triggered by the \texttt{main} function. KLEE could reproduce only 41 of the same exploits. 
\section{Greybox Fuzzing with Munch}\label{sec:munch}
Now that we've shown that isolating functions can lead to discovering more vulnerabilities, we will present a method to increase function coverage without deliberating symbolic execution at multiple entry points. 

The second use of sonar-search that we will highlight in this presentation is in the design of Munch \cite{ognawala2018munch}, our framework for hybrid fuzzing and symbolic execution. Munch tries to maximize \emph{function coverage} in C programs by adaptively switching between fuzzing and symbolic execution whenever one technique saturates in terms of function coverage. The motivation behind development of this tool was an observation that while symbolic execution is good at covering nodes of call-graph closer to the entry point of a program, coverage is, unfortunately, low for deeper lying functions. The reason for this may be path-explosion in the early sections of the programs or large constraints systems before calling deep-lying parsing functions. Fuzzing, on the other hand, while better at in-depth function coverage, is not as potent as symbolic execution in diversifying function coverage in the shallow regions of the call-graph.

Munch is an adaptive tool that operates in two modes of operations -- FS (Fuzzing+Symbolic execution, in that order) and SF (Symbolic execution+Fuzzing, in that order). In FS mode, the program is fuzzed (with AFL \cite{afl}) for a fixed amount of time. Then, the program is symbolically executed with sonar-search to target the functions that were not covered by the fuzzer. SF mode starts with symbolic execution (KLEE default strategy -- not sonar-search) and is followed by fuzzing with seed inputs that are generated by symbolic execution. 

Our evaluation results \cite{ognawala2018munch} on 9 open-source programs show that Munch performs better than symbolic execution or fuzzing alone, both, in terms of overall function coverage and function coverage at different depths of a program's call-graph. 

\section{Severity Assessment with Compositional Analysis}\label{sec:severity-assessment}
Having discussed methods to find more vulnerabilities with increased function coverage we now turn our attention to false-positives and how to mitigate them, empirically. 

We used sonar-search for severity assessment of vulnerabilities discovered using Macke and Munch. Essentially, this is a helper technique to increase the effectiveness of any compositional or hybrid technique involving symbolic execution. Severity assessment using compositional analysis was first introduced in \cite{ognawala2016macke} as an additional step but was further developed and improved by us. In the current state of our assessment framework, we predicted CVSS3 base score values \cite{mell2006common} for vulnerabilities discovered using Macke and Munch. For training our prediction model, we collected several existing vulnerabilities from the National Vulnerability Database (NVD) along with their CVSS3 scores. For all these vulnerabilities, we collected the affected versions of respective programs to analyze them using Macke. After analyzing them with Macke, we calculated certain heuristics (impact factors) for the affected \emph{functions} that contain the discovered vulnerabilities. In addition to other graph-based impact factors (e.g.\ function centrality, distance to program entry point etc.) we also included impact factors such as \emph{``length of longest error chain''}, which represents the number of interacting functions for which the same vulnerability (root cause) is exposed. A large value for such a chain may point to an inadvertent programming or oversight error, leading to a vulnerability. Using the ground truth and the corresponding impact factors, we were able to predict CVSS3 base score values for vulnerabilities that did not exist in NVD. 

Fortunately, we were able to reproduce all existing vulnerabilities in the NVD database with Macke. We evaluated the effectiveness of the severity assessment tool \cite{severityassessment} by presenting the results of 21 programs to experts and graduate students in the field of secure software engineering. The feedback about was largely positive, except for a few vulnerabilities which were harder for the experts to assess manually. 
\section{Conclusion}\label{sec:conclusion}
In this presentation, we aim to present the technical details of sonar-search, our targeted path-search implementation in KLEE, an equivalent of which did not exist in the erstwhile state-of-the-art. With three use-cases for targeted symbolic execution, we will, then, show the effectiveness of sonar-search in the context of compositional symbolic execution, greybox fuzzing and severity assessment of discovered vulnerabilities. 

\bibliographystyle{plainnat}
\bibliography{literature}

\begin{thebibliography}{8}
\providecommand{\natexlab}[1]{#1}
\providecommand{\url}[1]{\texttt{#1}}
\expandafter\ifx\csname urlstyle\endcsname\relax
  \providecommand{\doi}[1]{doi: #1}\else
  \providecommand{\doi}{doi: \begingroup \urlstyle{rm}\Url}\fi

\bibitem[afl()]{afl}
American fuzzy lop (afl).
\newblock \url{http://lcamtuf.coredump.cx/afl/}.
\newblock Accessed: 2018-01-09.

\bibitem[kle()]{klee22repo}
Klee22 source code repository.
\newblock \url{https://github.com/tum-i22/klee22/tree/sonar}.
\newblock Accessed: 2018-01-03.

\bibitem[sev()]{severityassessment}
Callgraph severity assessment tool.
\newblock \url{https://vmpretschner18.informatik.tu-muenchen.de/}.
\newblock Accessed: 2018-01-09.

\bibitem[Ma et~al.(2011)Ma, Yit~Phang, Foster, and Hicks]{ma2011directed}
Kin-Keung Ma, Khoo Yit~Phang, Jeffrey Foster, and Michael Hicks.
\newblock Directed symbolic execution.
\newblock \emph{Static Analysis}, 2011.

\bibitem[Mell et~al.(2006)Mell, Scarfone, and Romanosky]{mell2006common}
P.~Mell, K.~Scarfone, and S.~Romanosky.
\newblock Common vulnerability scoring system.
\newblock \emph{IEEE Security \& Privacy}, 2006.

\bibitem[Ognawala et~al.(2016)Ognawala, Ochoa, Pretschner, and
  Limmer]{ognawala2016macke}
S.~Ognawala, M.~Ochoa, A.~Pretschner, and T.~Limmer.
\newblock Macke: Compositional analysis of low-level vulnerabilities with
  symbolic execution.
\newblock In \emph{31st IEEE/ACM International Conference on Automated Software
  Engineering (ASE)}, 2016.

\bibitem[Ognawala et~al.(2018)Ognawala, Hutzelmann, Psallida, and
  Pretschner]{ognawala2018munch}
S.~Ognawala, T.~Hutzelmann, E.~Psallida, and A.~Pretschner.
\newblock Improving function coverage with munch: A hybrid fuzzing and directed
  symbolic execution approach.
\newblock In \emph{33rd ACM/SIGAPP Symposium on Applied Computing (SAC)}, page
  To appear, 2018.

\bibitem[Pretschner(2001)]{pretschner2001classical}
A.~Pretschner.
\newblock Classical search strategies for test case generation with constraint
  logic programming.
\newblock In \emph{Formal Approaches to Testing of Software}, 2001.

\end{thebibliography}
\end{document}